\documentclass[preprint]{aastex}

\usepackage{breqn}
\usepackage{amsmath}
 
\shorttitle{Temperature - dependent propagating disturbances }
\shortauthors{Uritsky et al.}

\begin{document}

\title{Measuring temperature - dependent propagating disturbances in coronal fan loops using multiple SDO/AIA channels and surfing transform technique}

\author{ Vadim M. Uritsky (1,2), Joseph M. Davila (2), Nicholeen M. Viall (2), and Leon Ofman (1,2) }

\affil{(1) Catholic University of America at NASA Goddard Space Flight Center, Greenbelt, MD 20771 USA}
\affil{(2) NASA Goddard Space Flight Center, Greenbelt, MD 20771 USA}






\begin{abstract}

A set of co-aligned high resolution images from the Atmospheric Imaging Assembly (AIA) on board the Solar Dynamics Observatory (SDO) is used to investigate propagating disturbances (PDs) in warm fan loops at the periphery of a non-flaring active region NOAA AR 11082. To measure PD speeds at multiple coronal temperatures, a new data analysis methodology is proposed enabling quantitative description of subvisual coronal motions with low signal-to-noise ratios of the order of $ 0.1 \%$. The technique operates with a set of one-dimensional ``surfing'' signals extracted from position-time plots of several AIA channels through a modified version of Radon transform. The signals are used to evaluate a two-dimensional power spectral density distribution in the frequency - velocity space which exhibits a resonance in the presence of quasi-periodic PDs. By applying this analysis to the same fan loop structures observed in several AIA channels, we found that the traveling velocity of PDs increases with the temperature of the coronal plasma following the square root dependence predicted for the slow mode magneto-acoustic wave which seems to be the dominating wave mode in the studied loop structures. 
This result extends recent observations by \citet{kiddie12} to a more general class of fan loop systems not associated with sunspots and demonstrating consistent slow mode activity in up to four AIA channels. 
 
\end{abstract}


\section{Introduction}
\label{sec:intro}

Propagating disturbances (PDs) are systematic translational motions of spatially localized emission enhancements in solar images.
Accurate identification of propagation speeds of PDs traveling along coronal loops is of critical importance for clarifying the underlying physical mechanism of this phenomenon. 

\citet{ofman97} and \citet{deforest98}, using data from respectively SOHO/UVCS and SOHO/EIT instruments, reported first observations of intensity perturbations along coronal plumes consistent with the slow magneto-acoustic wave. Signatures of slow mode waves have later been detected in both closed and open loop geometries using a variety of instruments, data analysis tools and simulation techniques \citep{demoortel00, nakariakov00, ofman99, ofman00, marsh09,prasad12, kiddie12}. The onset of Hinode/EIS era made the topic more controversial. Some authors found a clear in-phase relation between Doppler shift and intensity oscillations which could be attributed to a slow wave propagating upward from the transition region into the corona \citep{wang09, wang09a}. Others reported time-variable blueward asymmetry in spectroscopic EIS observations suggesting that the observed PDs can be caused by quasi-periodic plasma upflows and therefore the wave interpretation is not unique \citep{depontieu10, tian11a, tian11b, warren11, ugarte_urra11, mcintosh12, su12}. Some of these upflows may be coronal counterparts of type II spicules ejecting hot material from the chromosphere \citep{depontieu07a, depontieu11}. 

Flows and waves are not necessarily mutually exclusive. Quite often, they can coexist inside the same coronal loop, with upward plasma flows at footpoints driving slow wave activity at higher altitudes \citep{nishizuka11, ofman12}. In some cases, upward plasma flows are accompanied with downward flows reflecting cooling plasma dynamics \citep{kamio11, young12, mcintosh12}, making observational classification of PD signatures even more complicated. 

Analysis of temperature dependence of PD velocities adds an important piece of information to this picture by offering a potentially crucial test for the slow mode waves in coronal loops. If the PDs are caused by a slow magneto-acoustic wave, as opposed to a bulk plasma motion, their propagation velocities should be controlled by the local sound speed which is proportional to the square root of the plasma temperature. This dependence can be verified by comparing PD speeds in several bands of the extreme ultraviolet coronal emission. Since the relative amplitude of PDs tends to be low, determination of their speeds over a range of temperatures relies on data analysis algorithms enhancing the signal-to-noise ratios.

Until now, several types of techniques have been developed to reach this goal. The initial signal enhancements is commonly performed by using running-difference image sequences converted into time-differenced position time plots \citep{demoortel00}. Applying a moving average over a few consequent image frames helps achieve an acceptable signal-to-noise ratio. Some of the spatial resolution is usually also sacrificed to obtain a satisfactory confidence level. The velocity of the PD fronts can be roughly estimated using a simple visual inspection of the position-time plots, or by applying combined semi-automatic techniques, for instance by identifying the timings of peaks at each spatial position along the PD front, and fitting them using a linear regression model \citep{prasad12}. Any method relying on manual PD processing introduces a certain degree of subjectivity into the results. More elaborate techniques recruit a correlation analysis of PD signatures conducted independently of the observer. By measuring the cross-correlation functions between PD signals at each spatial position, it is possible to evaluate the velocity from the time lag ensuring the highest correlation for a given spatial separation \citep{kiddie12, tian11a, mcintosh12}. The apparent PD speeds can also be estimated by approximating the position-time plot with a propagating harmonic wave function characterized using a variety of fitting techniques as described by \citet{yuan12}. A single predefined propagating function described by constant values of phase speed, frequency, and phase shift, limit the range of applicability of this approach.

To enable a more accurate automated investigation of temperature-dependent wave fronts, this paper presents a new method for quantifying quasi-periodic PDs in high-resolution solar images based on the so-called surfing signals \citep{uritsky09} which are obtained by integrating a position - time plot along a varying spatiotemporal direction spanning a continuum of possible propagation speeds. This calculation, which we refer to as the { \it surfing transform } (ST), leads to an efficient self-averaging of random fluctuations preserving the initial time resolution of the image sequence, without distorting the frequency structure of the PDs. A two-dimensional (2D) Fourier power spectrum of the ST makes it possible to detect low-amplitude PD events as double resonances in the frequency - velocity space. We derive an analytical fit for the velocity resonance which can be used for measuring the scale  length, the amplitude, and the propagation speed of the disturbances. 

The developed technique makes it possible to quantify counter-streaming signals traveling simultaneously in opposite directions within the same loop structure. Our method does not rely on a particular wave form of the PD signal and can be applied to both harmonic and nonlinear propagating features. It can also be applied to repetitive propagating oscillations covering a wide range of frequencies without a well-defined characteristic periodicity. The primary limitation of the method is that it looses spatial information of the disturbance converting a spatio-temporal PD pattern into one-dimension temporal signals. In its present version, the transform is intended to investigate averaged PD characteristics and their time variability, but not detailed spatial inhomogeneities associated with propagation besides first-order damping effects incorporated into the applied fit.

The ST algorithm is applied to multi-temperature coronal images obtained from the Atmospheric Imaging Assembly (AIA) on board the Solar Dynamics Observatory (SDO). We study a fan loop structure of a non-flaring solar active region NOAA AR 11082 exhibiting propagating emission oscillation in several AIA channels, some of them with rather high noise levels. The results show a clear square root dependence of the PD propagation speed on the loop temperature consistent with the slow wave model. 

\section{Methods and data}
\label{sec:method}

\subsection{Surfing transform technique}

\begin{figure}[htbp]
\includegraphics*[width=8.0 cm]{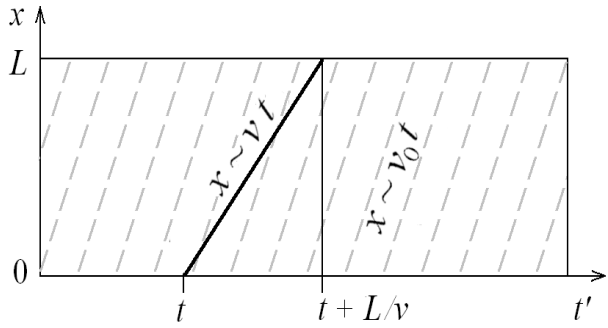}
\caption{\label{fig0} Schematic drawing illustrating the extraction of the ST signal from a position-time plot. Dashed lines show the fronts of a periodic PD described with phase speed $v_0$, solid line is the averaging path defined by the surfing speed $v$ and the starting time $t$, $L$ is the size of the studied region. }
\end{figure}

We define the ST as a time-dependent mean $a_S(t,v)$ of the position-time plot $a(x,t)$ computed along the straight line running through the point $x=0$ with the slope $v$ as shown in Figure \ref{fig0} :
\begin{equation}\label{eq1}
a_S(t,v) = \frac{v}{L} \int\limits_t^{t+ L/v } a(x=x_S, t') dt'.
\end{equation}
Here, $t$ is the starting time, $x_S = v\times(t' - t)$  is the averaging path, $L$ is the propagated distance, and $v$ is the surfing velocity. The calculation of the ST is conceptually close to the 2D Radon and Hough transforms (see e.g. \citet{jones09} and references therein); however, its application is different. Instead of using the integral transform (\ref{eq1}) for enhancing position-time plots, we focus on its resonance behavior.

If $a(x,t)$ is a periodic disturbance propagating with speed $v_0$, the dynamic range of $a_S$ maximizes at $v = v_0$. Consider ST of a one-dimensional harmonic wave oscillation :
\begin{equation}\label{eq2}
\begin{split}
a_S\left( t,v \right)  =  \frac{v}{L} \int\limits_{t}^{t + L/v} a_0 \, \mbox{cos} \left( k_0 v \times \left( t' - t \right) - \omega_0 t'  \right) dt' \\
  =  \frac{a_0 v}{L (k_0 v - \omega_0 ) } \left\{ \mbox{sin} \left( \left[k_0 v - \omega_0 \right]\left[ t+ L/v \right] - k_0 v t \right) -  \mbox{sin} \left( \left[ k_0 v -\omega_0 \right] t  - k_0 v t \right)  \right\} \\
  =  a_0 \, \frac{\mbox{sin} (\xi) }{ \xi} \,  \mbox{cos} \left( \omega_0 t + \xi \right) \,\,\,\, \mbox{with} \,\,\, \xi = \frac{L (k_0 v - \omega_0)}{2 v}, 
\end{split}
\end{equation}
in which we used the identity $\mbox{sin}(\alpha) - \mbox{sin}(\beta) = 2 \, \mbox{sin}( \frac{\alpha-\beta}{2} ) \mbox{cos}(\frac{\alpha+\beta}{2})$. In the dispersionless case, the phase speed $v_0 = \omega_0 / k_0$ and therefore the velocity detuning parameter $ \xi = L \omega_0 ( v - v_0)/(2 v v_0)$. As $v$ approaches $v_0$, $\xi$ tends to zero making $a_S(t,v)$ converge to $a(x=0,t)$. In this limit, the averaging paths in (\ref{eq1}) become constant phase lines yielding the largest possible ST amplitude given by $a_0$. 

Since (\ref{eq1}) is a linear procedure, the above result also applies to a sum of multiple harmonic modes. To investigate more complex processes, we calculate the {\it ST spectrum} describing 2D distribution of power spectral density over a range of Fourier frequencies and surfing velocities:
\begin{equation}\label{eq3}
P_S(v, \! f)\! = \! \frac{v}{L} \left| \int\limits_{-\infty}^{\infty} \!\! \int\limits_{t}^{\,\,\,\, t + L/v} \!\!\!\!\! a(v\times(t' \! -t), t') e^{-i 2 \pi f t} dt' dt \right|^2,
\end{equation}
as well as the {\it velocity spectrogram} $P_S(v,f=f_0)$ for a fixed frequency $f_0$. The shape of the velocity spectrogram of a harmonic wave at $f_0=\omega_0/ (2 \pi)$ is given by the square of the $\mbox{sinc}(\xi)$ function (see Eq. \ref{eq2}) reaching its maximum at $v=v_0$:
\begin{equation}\label{eq5}
P_S = P_0 \,\, \mbox{sinc}^2 ( L \omega_0 ( v - v_0) /(2 v v_0))
\end{equation}
where $P_0$ is the height of the spectral peak. As the propagation distance $L$ and the frequency $\omega_0$ of the oscillation increase, the peak becomes narrower and the resonance measurements more precise. The increase of $v_0$ has the opposite effect leading to a broader $P_S(v)$ peak.

\subsection{Statistical testing}

Panels (a) through (e) of Figure \ref{fig1} show the application of the described technique to three superposed wave signals mixed with a correlated Brownian noise \citep{mandelbrot68} resulting in a 10$\%$ signal-to-noise ratio:
\begin{equation}\label{eq4}
a(x,t) = \sum_{i=1}^3 a_{0} \, \mbox{cos} \, (k_i x - \omega_i t) + \sigma \zeta_{x, t},
\end{equation}
in which $[\omega_1, \omega_2, \omega_3]/(2 \pi) = [5.00, 3.33, 3.33]$ mHz, $[ v_1, v_2, v_3 ] = [-100, 50, 120 ]$ km/s,  and $k_i = \omega_i / v_i$. The signal with negative velocity propagates in the negative $x$ direction. The surfing signals in panel (c) are extracted from the position-time plot (a) by tuning $v$ to the three wave speeds. The ST spectrum shown in panel (d) features three peaks corresponding to these waves. The velocity spectrograms representing horizontal cuts of the ST spectrum through the wave frequencies (two of which coincide) are provided on panel (e). For a more precise result, the spectrograms were averaged over the frequency bands shown on panel (d) by dashed horizontal lines. The spectrograms demonstrate surfing resonances at the expected $v$ values. The negative velocity peak is easily identifiable demonstrating the ability of the method to recognize PDs traveling downward and upward along coronal loops during the same observation interval. This capability constitutes one of the main advantages of the developed method compared to the techniques relying on cross-correlation analysis \citep{kiddie12} or nonlinear regression fits \citep{yuan12} using single-component wave models.

The phase speed of the slow mode waves in coronal loops is controlled by the local sound speed $c_s^2 = \gamma \, p / \rho \propto T$, where $\gamma$ is the adiabatic index, $p$ and $\rho$ are respectively the kinetic pressure and the mass density of the coronal plasma, and $T$ is the plasma temperature. Assuming that the wave propagates in a bundle of isothermal coronal loops excited by the same periodic process in the transition region but described by different values of $T$, we expect the PD velocities in these loops to scale as the square root of $T$ while the frequencies to be approximately the same. 

The velocity-temperature scaling $v_0 \propto T^{1/2}$ indicative of the slow mode can be tested based on ST analysis of different SDO AIA channels with distinct temperature peaks of sensitivity \citep{boerner12, lemen12}. Figure \ref{fig2} presents numerical simulation of such multitemperature measurement in the presence of additive Brownian noise. The first column shows position-time plots of harmonic wave oscillations modeling slow wave with $v_0=90$ km/s observed in the 171 \AA\ channel. The noise intensity increases from top to bottom. Similar $a(x,t)$ plots were constructed for 131 \AA , 193 \AA , and 211 \AA\ AIA channels, with $v_0$ satisfying the temperature scaling of the slow magneto-acoustic wave (respectively 68 km/s, 136 km/s, and 144 km/s). The second column shows the velocity spectrograms $P_S(v)$ of the channels computed at the base wave frequency $f_0=3.0$ mHz. Parameters $v_0$, $L$ and $P_0$ were adjusted to obtain the best least-square fits (\ref{eq5}) plotted with solid black lines. Solid and dashed vertical lines show the estimated and true $v_0$ values, correspondingly. It can be seen that our method yields reasonable velocity estimates for quite low signal-to-noise ratios. 

The last two rows of panels in Figure \ref{fig2} show the results of a null testing of our method using Brownian and white noise models without admixtures of periodic signals. In both cases, the ST spectrograms reveal no well-defined velocity peaks observed in the presence of PDs. 

\subsection{SDO AIA observations}

The described  methodology was applied to SDO AIA images of solar active region NOAA AR 11082 collected on 2010 June 19 during 4:00 - 10:00 UT with angular resolution $\sim$0.6 arcsec and the maximum available cadence time $\sim$12 s. The region was located in the northern hemisphere close to the disk center. It had a bipolar structure not associated with a sunspot, and showed no flaring activity above B-class for the studied time period. We use level 1.5 data from the cutout service. The images were derotated by the SHIFT\_IMG routine of the SolarSoft package, and co-aligned as described by \citet{viall11,viall12}. As can be seen from  Figure \ref{fig3}, NOAA AR 11082 has an extended system of fan loops. The PD activity in these loops was studied using 131 \AA , 171 \AA , 193 \AA , and 211 \AA\ channels. The remaining AIA channels showed insufficient count rates and were excluded from the analysis. 

Three of the studied AIA passbands (131 \AA\ , 193 \AA\ and 211 \AA\ ) are known to be described by multithermal response functions \citep{delzanna12, delzanna12a, delzanna12b} which require special attention. The 131 \AA\ channel reflecting the contribution from Fe VIII, Fe XX, and Fe XXIII lines has two distinct peaks of sensitivity at about 0.56 MK and 10.8 MK. The high-temperature peak is $\sim$2.5 times lower than the first peak \citep{boerner12, lemen12}. The significance of the high-temperature 131 \AA\ for our analysis is limited for two reasons. First, the studied non-flaring coronal region is relatively cool and most likely contains no $\sim 10^{7}$ K plasma. Second, even if such plasma exists, the expected slow mode speed at this temperature is beyond the range of measurable speeds. According to \citet{deforest98} and subsequent studies (see e.g. \citet{aschwanden06} for a review), the characteristic traveling velocity of acoustic waves in coronal structures observed in the approximately monothermal 171 \AA\ channel ($T\approx$ 0.9 MK) is of the order 100 km/s which translates into $\sim$350 km/s at $T=$10.8 MK. On the 24 s Nyquist time scale (two AIA sampling intervals), such wave would travel a distance $\sim$8.5 Mm comparable with the projected length for the studied loop region, and cannot be accurately detected by any method. Based on these arguments, we consider the high -temperature peak of the 131 \AA\ emission to be irrelevant to our analysis and focus on the main sensitivity peak of this channel.

The 193 \AA\ passband is also multithermal reflecting a strong contribution from Fe VIII, Fe IX, Fe XI and Fe VII lines, as well as a much weaker emission from lower-temperature lines dominated by O V and Fe VII \citep{delzanna11, kiddie12}. The main cool contribution to the 193 \AA\ passband comes from Fe VIII and Fe IX lines. As shown by \citet{kiddie12}, it can reach up to 40$\%$ of the total emission in this channel in sunspot regions and in that case cannot be ignored. According to the same study, the significance of the cool 193 \AA\ component is much lower in non-sunspot regions such as NOAA AR 11082 where it can be responsible for about 15-25 $\%$ of the observed emission. Due to smallness of this contribution we do not address it directly keeping in mind that the estimated speeds can be somewhat slower than what could be expected for the main 193 \AA\ sensitivity peak centered at 1.6 MK. As shown in the next Section, such bias does exist and it is small compared to the observed temperature dependence. 

The low-temperature emission of the 211 \AA\ passband is generally by a factor of 2 smaller than the corresponding contribution of the 193 \AA\ passband \citep{boerner12}. Taking this fact into account, we associate the temperature of the 211 \AA\ channel with its main response function peak around 1.8 MK.

\section{ST analysis results}

\subsection{PD activity in selected fan loop region}

We found that many of the studied fan loops sectors carry quasi-periodic PDs during the entire 6 hour interval of observations. The disturbances travel in predominantly upward direction away from the base of the corona, with the periodicity of 4 - 8 min an the apparent propagation speed $\sim 40$ to $\sim 180$ km/s. Here we focus the fan loop sector highlighted by dashed white lines in Figure \ref{fig3}(a) which showed the most stable PD activity. 


Figure \ref{fig4} shows the results of the ST analysis of the chosen sector. The position-time plots shown on left panels exhibit PD patterns. To enhance the PD signals, the plots were subject to temporal differencing with the 60 s time lag. The pairs of straight lines added to each panel visualize measured PD velocities, with the interval between the lines showing the characteristic relaxation time $f_0^{-1} \sim 310$ s of PD fronts given by the highest resolvable harmonic of the ST spectrum in the least noisy 171 \AA\ channel. The ST spectra of 171 \AA\ emission typically contain several harmonics caused by the intermittent structure of the studied signals. The highest harmonic used hear represents the characteristic duration of single intensity perturbations and is the most relevant to the velocity analysis of the individual PD fronts. 

The velocity spectrograms fitted with Eq. (\ref{eq5}) are presented on the central panels. The obtained velocity estimates (marked by solid vertical lines) are approximately consistent with the slow mode speeds (dashed lines) predicted based on the measured 171 \AA\ channel speed. The resonance peaks are broader for larger $v_0$ values, in agreement with the analytical dependence (\ref{eq5}). 


The estimated scale length $L$ varies between $\sim$ 11 Mm at 131 \AA\ and 20-21 Mm at 171 \AA\ and 193 \AA\ wavelengths. This behavior could be a mixture of several factors, such as e.g. different levels of noise in the AIA channels, temperature - dependent scale height of sound waves in radially diverging coronal magnetic field (e.g., \citet{torkelsson98}), and viscous dissipation of wave energy leading to pronounced altitudinal stratification of the slow mode amplitude \citep{ofman00}. 

The right column of panels in Figure \ref{fig4} shows surfing signals obtained from the respective $a(x,t)$ plots using the estimated PD speeds. The shape of the signals confirms the periodic nature of the observed activity. The apparent cross-channel coherence of the signals which represent temporal oscillations at the base ($x=0$) of the studied wave region suggests that the wave fronts in the studied multi-temperature loops were excited nearly simultaneously at the base of the corona, possibly by a global $p$-mode oscillation. The coherence is the most pronounced for the strongest wave fronts. For instance, the two intense wave fronts marked by dashed vertical lines are seen at approximately the same times in all AIA channels. We conducted a separate analysis of this pair of wave fronts withing a narrowed loop sector (solid white boundary in Figure \ref{fig3}(a)) where the fronts were the most intense. The position-time plots of the fronts were replicated several times to enable the application of our ST algorithms to episodic PD activity. 

Table \ref{table1} reports the velocities of the selected replicated fronts normalized by the velocity in the 171 \AA\ channel, along with the normalized velocities obtained for the entire time interval shown in Figure \ref{fig4}. The errors are the propagated standard deviations from the nonlinear least-square fit (\ref{eq5}) with the Gaussian weighting. For reference purpose, the last column of table provides the velocity ratios for the sound wave. Comparing velocity ratios has the advantage of eliminating the uncertainty associated with projection effects and the unknown value of $\gamma$ which can vary between 1 and 5/3 \citep{vandoors11}. It can be seen that the velocity ratios measured by our method are fairly close to the prediction for the slow mode wave. The agreement with the theory is somewhat better for the selected intense fronts due to their higher signal-to-noise ratio. 

Our analysis revealed no signatures of downwardly propagating PDs in the studied loop coexisting with the temperature-dependent upward motion. Figure \ref{fig5} presents bidirectional velocity spectrograms which were integrated over a frequency range 0.002 - 0.020 Hz to take into consideration broadband emission oscillations. This range includes the PD activity  described above as well as other possible motions associated with flows and waves. It can be seen that the spectral power in the 171, 193 and 211 \AA\ passbands associated with negative propagation direction is by a factor 4-5 smaller compared to the power in the positive (upward) direction and shows no distinct maxima at $v_0 < 0$. The much lower count rates of the 131 \AA\ channel leave a possibility of an unresolved downward motion in this channel. Overall, Figure \ref{fig5} suggests that even if an undetected downward motion is present its intensity is negligibly small compared to that of upwardly traveling disturbances.  


\subsection{Distribution of PDs across other fan loops}

Using the same methodology, we studied spatial distribution of PD activity across the loops belonging to the same active region. The fan loop system of NOAA AR 11082 was divided into 50 narrow sectors diverging radially from the center of the fan loop structure and characterized by equal angular ($\sim 2.3^{\circ}$) and radial ($\sim 17.3$ Mm) sizes. 
The sectors were approximately parallel to the local magnetic field traced by the fan loops. For each sector and the AIA channel, we constructed a position-time plot describing spatial distribution of the AIA intensity along the sector as a function of time during the same interval as the investigated in Figure \ref{fig4}. To enhance periodic signals, the position-time plots were subject to 60-second temporal differencing as before. 

The results of the survey have shown upward PD activity is ubiquitously present throughout the studied fan system as shown in Figure \ref{fig6}(a). The results of the velocity measurements in different sectors are provided in Table \ref{table2}. These measurements confirm the tendency of the PD velocity to increase with the temperature of the coronal plasma. Most of the sectors exhibited sufficiently strong PDs in at least two AIA channels making it possible to compute the velocity ratios in the individual fan loop  sectors and compare the estimated values with the slow wave mode prediction. Panel (b) of Figure \ref{fig6} presents the results of this verification. Here, the velocity ratios estimated for several loop sectors are plotted using symbols of various pattern as shown in the plot legend. The plotted values were aggregated into groups of five adjacent sectors to improve statistical accuracy. The dash-dotted line represents the average between these values for each temperature value. 

The observed temperature scaling is in a reasonable agreement with the theoretical square-root dependence plotted with the solid black line. This tendency is statistically significant for 193 \AA\ and 131 \AA\ velocity ratios but not for the 211 \AA\ ratio due to larger propagation velocities observed in this channel (Table \ref{table2}) yielding a broader $P_S$ peak (see Eq. \ref{eq5}) leading to a greater statistical uncertainty. It is interesting to note that the 193 \AA\ velocity ratios are systematically below the theoretical slow mode dependence. This bias is likely to reflect the cool emission component present in this passband. The 131 \AA\ velocity ratios are also systematically smaller than the predicted value, but most probably due to the high level of noise in this channel leading to false $P_S$ peaks at $v_0 \sim$ 40-50 km/s as suggested by statistical tests presented in Figure \ref{fig3}. Such peaks reflecting the correlation structure of non-propagating fluctuations can interfere with the actual velocity resonances causing an asymmetric broadening of $P_S$ peaks toward lower PD speeds and yielding underestimated $v_0$ values. Some of the bias can also be caused by the contribution from the cool branch of the 131 \AA\ temperature response \citep{tian11a, mcintosh12}. The magnitude of these effects, which could in principle be removed using a more sophisticated fitting function, is nevertheless smaller than the observed square-root temperature trend. 

\subsection{Relationship with flows}

Our results strongly suggest that the studied fan loop system is dominated by upwardly moving disturbances which are consistent with the slow magneto-acoustic wave model. The unavailability of high-cadence spectroscopic observations for these events makes it impossible to verify this interpretation using Doppler measurements as was done by other authors, see e.g. \citet{wang09, wang09a}. However, the obtained evidence presents an explicit case in favor of the slow mode waves in the studies solar region which is hard to ignore. 

It should be clarified that the observed temperature dependence of PD velocity cannot not be associated with a superposition of downward and upward bulk motions considered in some previous works. A number of recent studies reported unambiguous signatures of such flows using conjugate coronal imaging and spectroscopic observations. Some of these works are already mentioned in the Introduction. In particular, \citet{depontieu10} have identified ubiquitous and faint upflows with characteristic speeds 50-150 km/s in coronal loops associated with plage regions. These flows caused blueward asymmetries of spectral line profiles in footpoint regions of coronal loops and could be related to high-temperature plasma jets originated emitted from the chromosphere. The hot plasma upflows can be accompanied by downflows of radiatevily cooling material closing the mass cycle of the coupled chromosphere - corona system \citep{mcintosh12}. The presence of counterstreaming flows complicates the interpretation of line profiles which can exhibit a net Doppler shift asymmetry suggestive of an apparent temperature-dependent plasma speed. 

We note, however,  that such contraflows would be resolved by the ST technique were they present in the fan loop system studied here. As shown in Section \ref{sec:method}, the ST analysis is able to differentiate between counter-propagating disturbances and would selectively detect both types of PD if they were simultaneously present at either different or the same coronal altitudes. However, no such periodic contraflow events \citep{mcintosh12} were found in the studied fan loop regions. The 131 \AA\ passband which exhibits a roughly symmetric ST velocity spectrogram (Figure \ref{fig7}) has very low count rates and is likely to be dominated by noise. In some locations, we did observe occasional features descending towards the transition region such as the one shown in Figure \ref{fig7}. A closer inspection has shown that these features are caused by a proper motion of several closed loop filaments relative to the steady fan loop system. This effect is similar to the intensity oscillations associated with Alfv\'enic  waves observed by \citet{tian12a} causing loop displacement rather than density variations, and it cannot be attributed to plasma downflows.

Obviously, the existence of plasma flows can not be ruled out based on the imaging diagnostics alone. Based on our measurements, it is unlikely that bulk plasma flows play a leading role at the coronal altitudes covered by our analysis. The  observed emission dynamics indicates that (a) the motion occurs only in one direction and (b) it exhibits wave-like temperature scaling. However, there is a possibility that significant flow motions unresolved by our method exist near the footpoints of the studied fan loops. These motions could be synchronized by $p$-mode oscillations and be part of the global chromosphere-corona mass cycle as proposed by \citet{mcintosh12}. At the same time, they could excite slow mode and other magnetohydrodynamic waves in the upper coronal regions. In this scenario, the faster propagating upward flows would provide energy for the waves. In the open field line geometry, the waves can carry a sizable portion of this energy away from the transition region into the solar wind. A similar flow-wave interaction has been recently reproduced in magnetohydrodynamic simulations  of a coronal active region \citep{ofman12, wang13}. More experimental and theoretical studies will be needed to verify the validity of the scenario and its underlying physics.

\section{Conclusions}

We have presented a new methodology for measuring parameters of PDs in multi-temperature coronal images. The key element of the proposed approach is the analysis of surfing signals which demonstrate a resonance behavior in the presence of quasi-periodic PDs. The developed methodology has been applied to the analysis of PDs in coronal fan loops in AR 11082 across a range of temperatures covered by four SDO AIA channels. We found that traveling velocity of PDs propagating along the loops obeys rather accurately a square root temperature dependence predicted for slow mode magneto-acoustic waves which seems to be the dominating wave mode in the studied loop structures. 

Recently, \citet{kiddie12} have arrived at a similar conclusion based on a detailed analysis of PDs in coronal fan loops anchored at sunspots. Using three AIA channels, they have found the velocities of such PDs depend on the temperature as predicted for the slow wave mode. However, they reported no conclusive observations for PDs in non-sunspot locations. Our study extends their findings by showing that the temperature-dependent slow mode activity can be observed simultaneously in four AIA channels in a non-sunspot active region. It remains to be clarified whether or not this wave activity is driven by undetected fast plasma flows at the base of the corona as has been proposed by some authors \citep{depontieu10, tian11a, tian11b, warren11, ugarte_urra11, mcintosh12, su12, ofman12}.

The range of applicability of the presented technique involves a variety of more general tasks beyond the scope of this paper. For instance, it would be of interest to apply the technique to plume regions. Plumes were the first coronal structures where slow magnetosonic waves were identified (e.g., \citep{ofman97, deforest98, ofman99}. Recently, \citet{tian11c} have found no clear temperature dependence of PDs observed in plume-like structures rooted in various solar regions including polar and equatorial coronal holes as well as quiet Sun. The velocity measurements used in their study were based on visual inspection of position-time plots. The technique described in our paper can be used to obtain more accurate PD velocity estimates in such structures and to  test earlier conclusions regarding the presence of absence of the magneto-acoustic $v - T $ scaling in multithermal plumes.

\begin{figure*}[htbp]
\includegraphics*[width=7.0 cm]{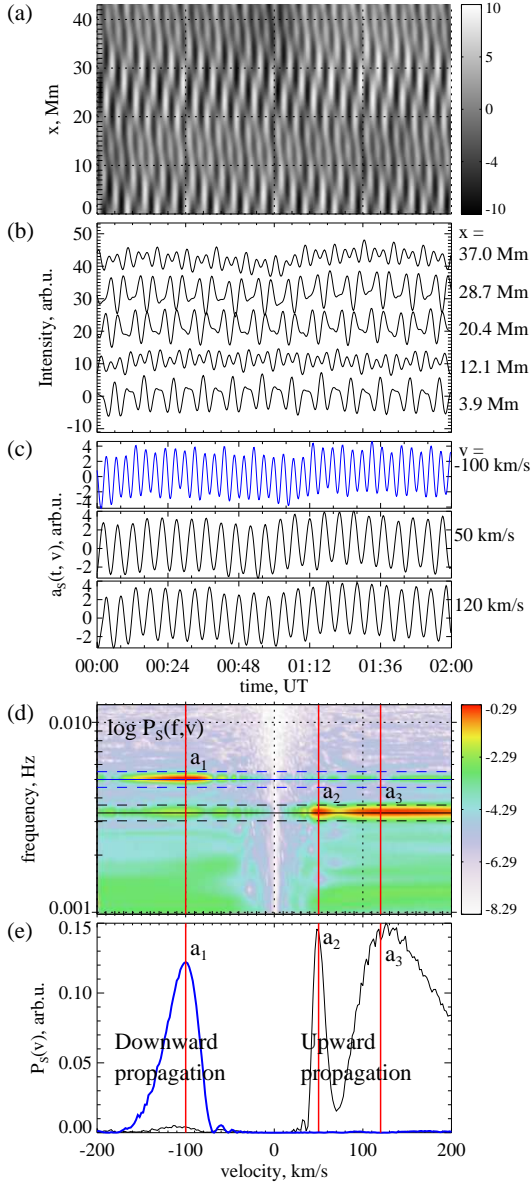}
\caption{\label{fig1} Testing the ST technique on upward and downward PDs. (a) Position-time plot of the simulated wave signal defined by Eq. (\ref{eq4}). (b) Stack plot of the signal at five selected $x$ positions. (c) Surfing signals obtained using three values of $v$ matching the phase speeds of the wave components. (d) ST spectrum computed using 400 ST signals covering the velocity range $v \in [-200, 200]$ km/s. (e) Velocity spectrograms representing horizontal sections of the ST spectrum at $f_0 = 5.00$ mHz (dotted blue line) and $f_0 = 3.3$ mHz (solid black line). Note that the method recognizes concurrent downward and upward motions in the same of different frequency bands.}
\end{figure*}

\begin{figure*}[htbp]
\includegraphics*[width=14 cm]{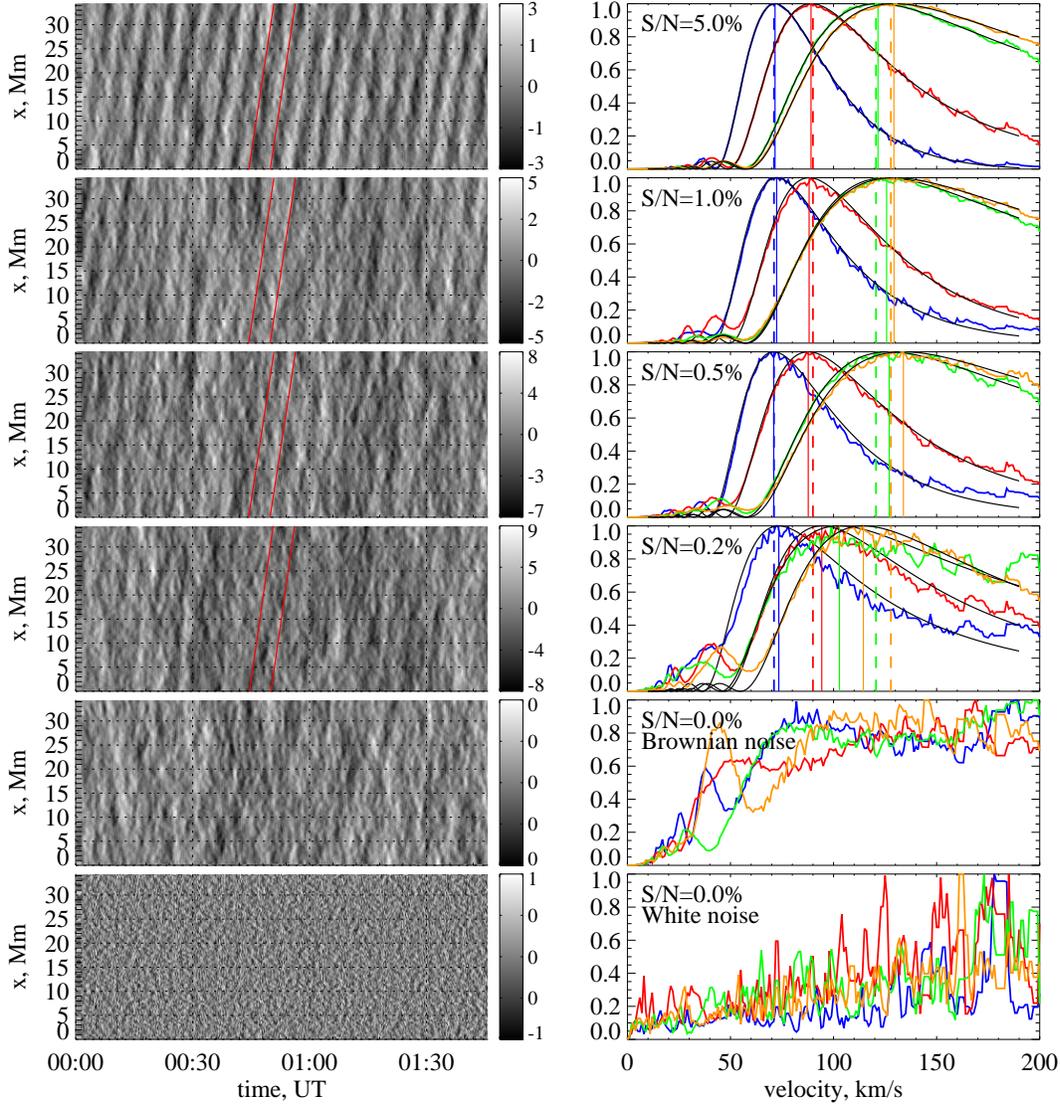} 
\caption{\label{fig2} ST analysis of simulated wave signals in four SDO AIA channels, as compared to no-propagating random noises. Left: time-differenced position-time plots of the 171 \AA\ channel. Right: velocity spectrograms obtained for each channel; the phase speed of the wave is adjusted to the temperature the channel assuming that the wave is slow magnetosonic. Color coding: 131 \AA\ - blue, 171 \AA\ - red, 193 \AA\ - green, 211 \AA\ - brown. Black curves are least-square fits using the theoretical model (\ref{eq5}), solid (dashed) vertical lines are the measured (predicted) propagation velocities for each channel. The amount of additive noise increases from top to bottom as shown by the provided power signal-to-noise ratios (S/N)/. The tilted red lines on the left panels show the phase speeds and wave periods measured at 171 \AA . The last two rows of panels corresponding to the cases of random fluctuations with Brownian and ``white'' spatial spectra showing no wave PD signatures on ST spectrograms, as expected. }
\end{figure*}

\begin{figure*}[htbp]

\hspace{0.3 cm} (a) \hspace{8.0 cm} (b)
\begin{center}

\includegraphics*[width=7.8 cm]{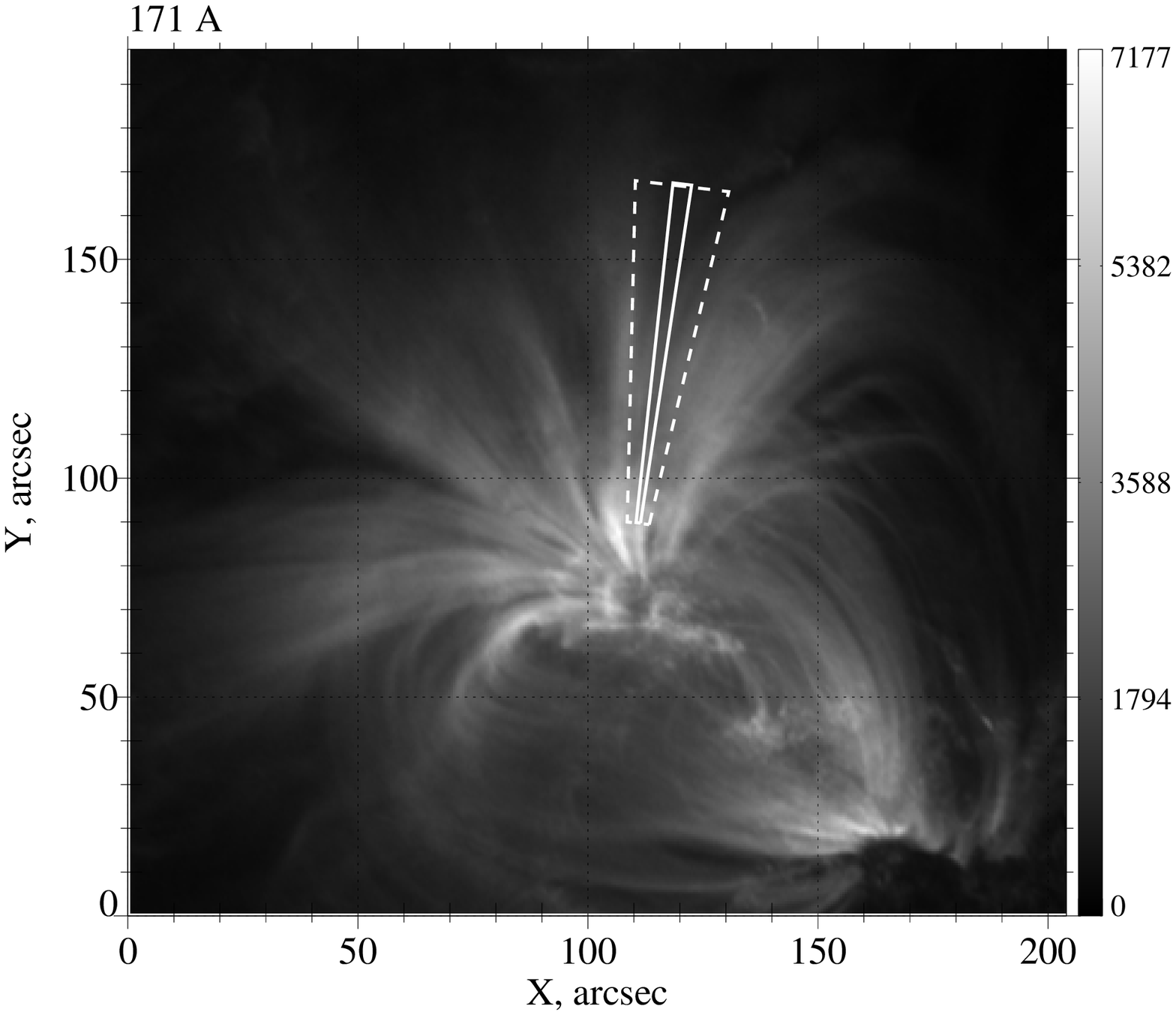} \includegraphics*[width=8.2 cm]{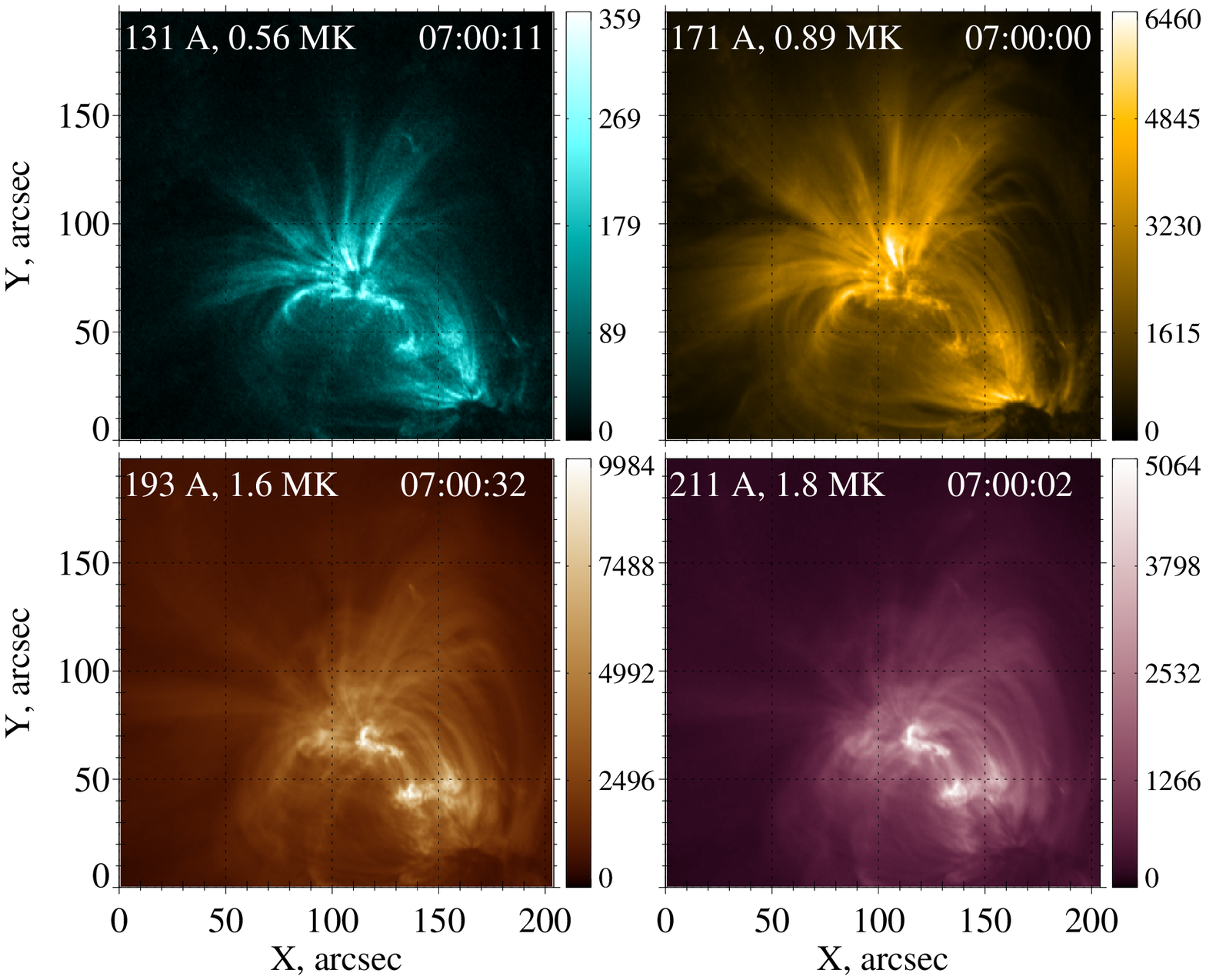} 

\end{center}

\caption{\label{fig3} (a) A snapshot of the active region NOAA AR 11082 as seen in the 171 \AA\ SDO AIA channel at 7:00:00 UT.  The dashed white polygon shows the domain of the PD activity studied in Figures \ref{fig4}-\ref{fig5}; the solid white line outlines the adjusted domain used to analyze two selected PDs reported in Table \ref{table1}.
(b) Multi-spectral AIA images of the same region obtained at around the same time; temperatures values are for the peaks of the response functions based on the 2013 release of the CHIANTI atomic database \citep{dere97, landi13}. }

\end{figure*}

\begin{figure*}[htbp]
\begin{center}
\includegraphics*[width=15 cm]{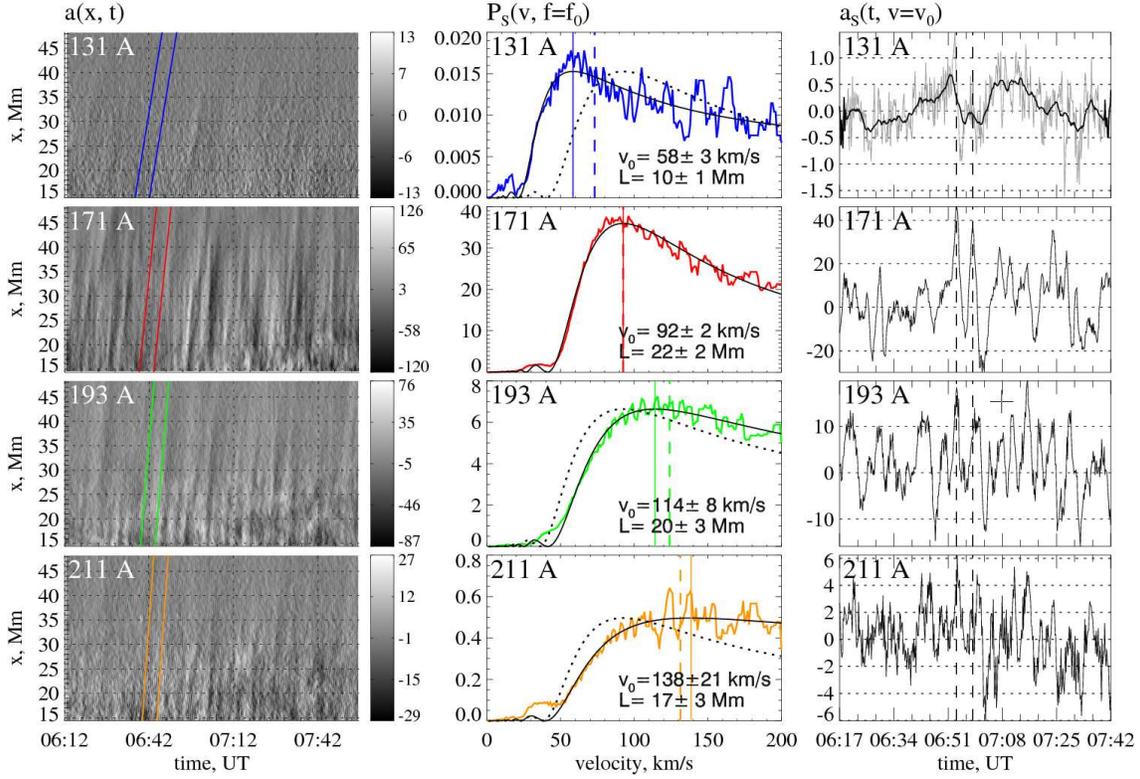} 
\end{center}
\caption{\label{fig4} Analysis of quasi-periodic PDs observed in the fan loop sector outlined by dashed white line in Figure  \ref{fig3}(a):  position-time plots (left), ST-based velocity spectrograms (center), and ST signals constructed using $v_0$ values measured for each passband. Black solid curves overplotted with velocity spectrograms are the best least-square fits based on Eq. (\ref{eq5}) yielding the estimates of the propagation velocity $v_0$ and the projected PD scale length $L$. Dotted lines show the shape of each spectrogram under the assumption of temperature-independent wave propagation at the 171 \AA\ phase speed. Solid vertical lines show the measured PD speeds; dashed vertical lines are the slow magnetosonic wave velocities predicted based on the 171 \AA\ measurement. Vertical lines on ST signal mark the timing of the two selected fronts in the 171 \AA\ channel discussed in the text. The most noisy 131 \AA\ channel is represented by the original (gray) and the low-pass filtered (black) ST signals. } 
\end{figure*}

\begin{figure}[htbp]
\includegraphics*[width=7 cm]{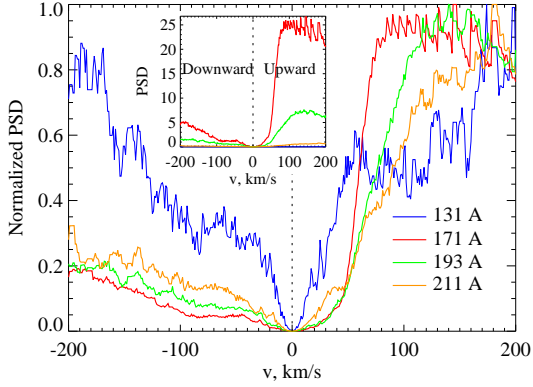} 
\caption{\label{fig5} Bidirectional velocity spectrograms obtained by integrating ST spectra over the frequency range 0.002 - 0.02 Hz (periods 50 - 500 s, or 0.83 - 8.3 min). The 171, 193 ans 211  \AA\ passbands show the prevalence of upwardly moving features. The 131 \AA\ spectrogram is approximately symmetric due to low count rates in this AIA channel as can be seen from the inset plot displaying non-normalized $P_S (v)$ plots. }
\end{figure}

\begin{figure}[htbp]
\hspace{0.2 cm} (a) \hspace{7.8 cm} (b)

\includegraphics*[width=7.0 cm]{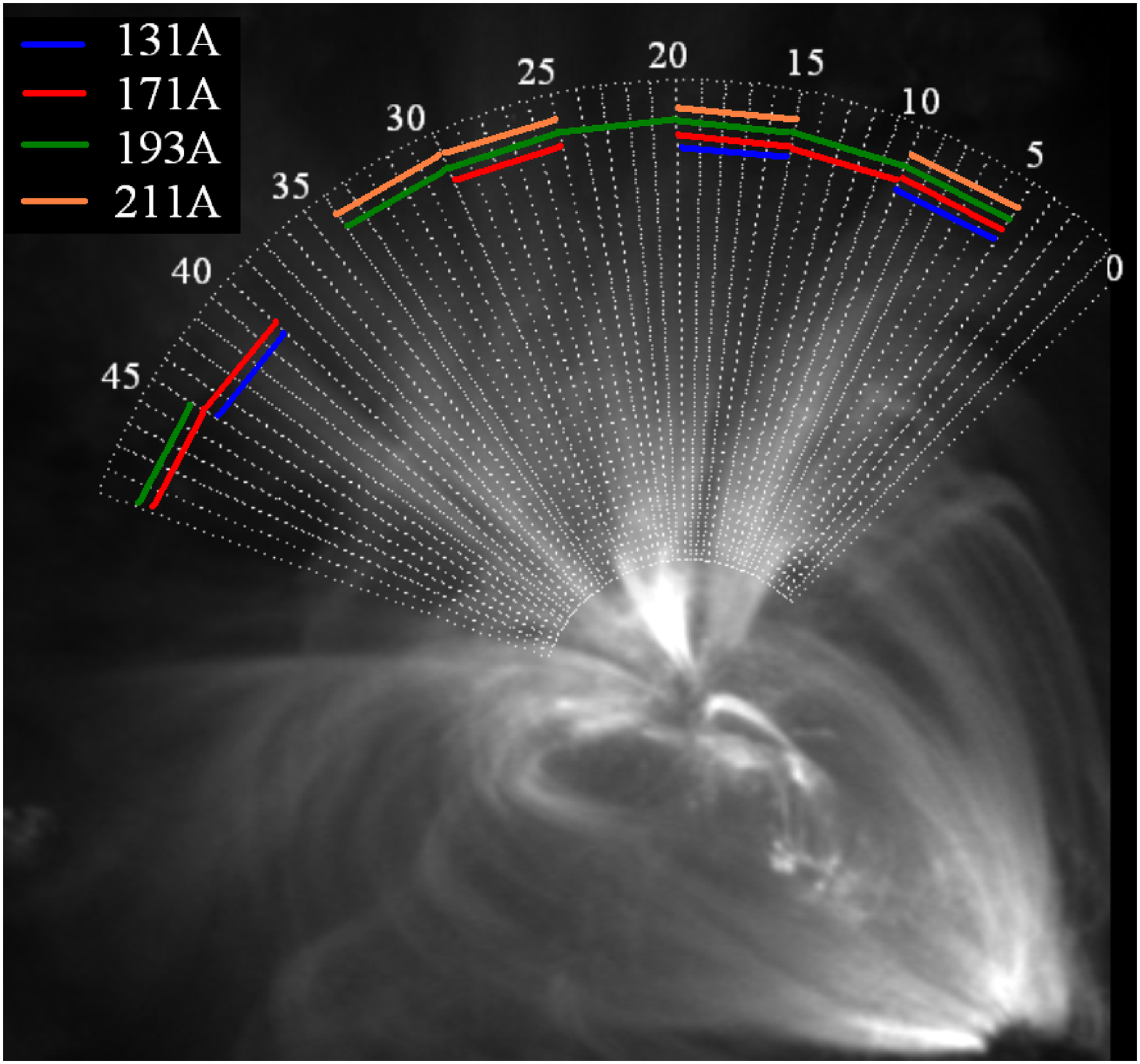} \includegraphics*[width=9 cm]{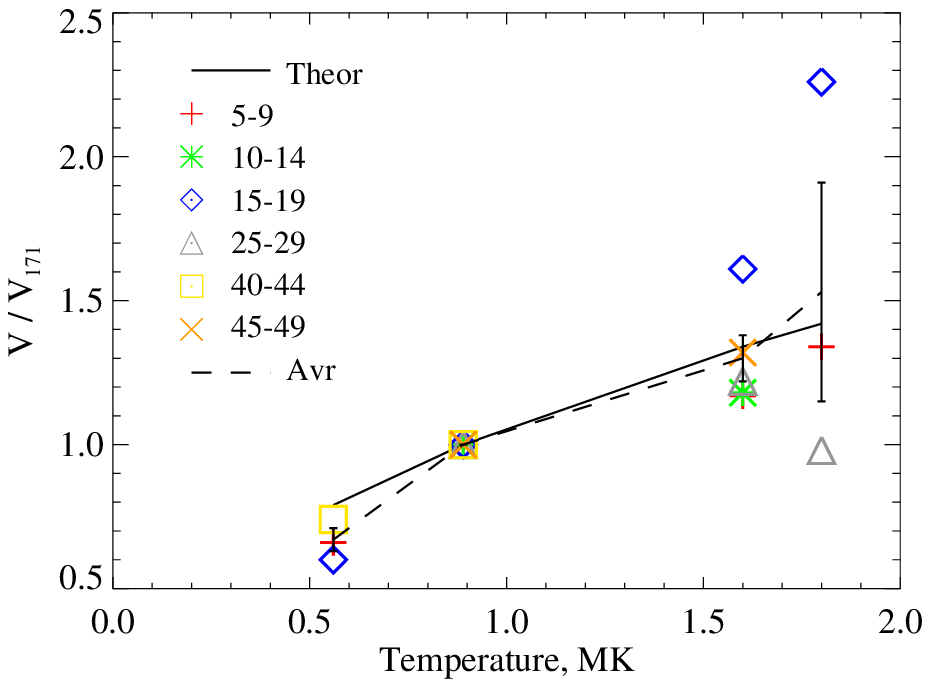} 
\caption{\label{fig6} . Results of the PD survey using the ST technique. (a) 171 \AA\ image divided into sectors. The color-coded bars show the presence of identifiable PD activity in the corresponding AIA channel. (b) Velocity ratios between the measured PD velocities and the velocity in the 171 \AA\ passband versus loop temperature given by maximum SDO AIA sensitivity in each channel. The theoretical square-root dependence expected for slow mode waves is plotted with black solid line; symbols show the results for the fan loop sectors indexed in panel (a) in which PDs were observed at one or more AIA wavelengths including 171 \AA\ . Dashed line represents the average velocity ratio characterizing different fan sectors, with vertical error bars showing plus-minus one standard error. }
\end{figure}

\begin{figure}[htbp]
\includegraphics*[width=7 cm]{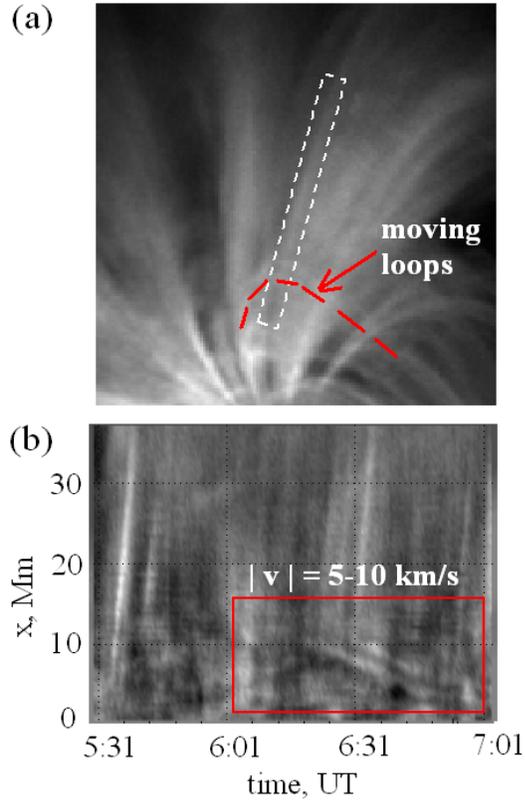} 
\caption{\label{fig7}. Example of apparent downward motion in the lower portion of the studied fan loop system coexisting with upward PDs at higher altitudes. Panel (a) shows the analyzed region outlined by a white dashed polygon; panel (b) is the detrended position-time plot representing this region. The downward PDs in this event are an artifact caused by the displacement of a set of descending closed loops marked with a dashed line on panel (a) which cross the fan loop filaments included in the virtual slit. } 
\end{figure}

In principle, ST analysis can be applied to any superpositions of multiple PD processes characterized by different velocities and/or frequencies as illustrated in Figure \ref{fig1}, including strongly nonlinear wave forms. It could be especially helpful for analyzing wave damping effects since the ST algorithm provides self-consistent evaluation of the amplitude and the scale length of the wave. Such measurements can conducted in various astrophysical contexts not limited to solar corona. 

\acknowledgments
The authors would like to thank James Klimchuk and Tongjian Wang for useful discussions.  V. U. was supported by the NASA Grant NNG11PL10A 670.002 through the CUA's Institute for Astrophysics and Computational Sciences. L. O. was supported by NASA grant NNX12AB34G.  N. V. thanks Harry Warren for help with the derotation methodology. 

\clearpage

\clearpage


\begin{table}
\begin{center}
\caption{\label{table1} Velocity ratios measured for the entire wave event and two selected fronts, as compared to the predicted sound speed ratios. See text for details. }
\begin{tabular}{lccc}
\hline
Wavelength & All fronts &  Selected fronts & Predicted  \\
\hline
131 \AA\ &	0.63 $\pm$ 0.03		&	0.86 $\pm$ 0.04		&	0.79		\\

171 \AA\ &	1 &	1	&	1	\\

193 \AA\ &	1.23 $\pm$ 0.08		&	1.31 $\pm$ 0.06		&	1.34		\\ 

211 \AA\ &	1.50	$\pm$ 0.23	&	1.50	$\pm$ 0.15	&	1.42		\\ 

\hline
\end{tabular}
\end{center}
\end{table}


\begin{table}
\begin{center}
\caption{\label{table2} Parameters of PD activity in the fan loop sectors shown in Figure \ref{fig5}. }
\begin{tabular}{l|c|cccc}
\hline
slice index & $f$, mHz & & $v_0$, km/s &  &  \\
 & & 131 \AA\ & 171 \AA\  & 193 \AA\  & 211 \AA\ \\
\hline
0 - 4   &      -      &     -      &     -      &      -       &     -        \\
5 - 9   & 1.42 - 3.19 & 47 $\pm$ 3 & 71 $\pm$ 5 & 83 $\pm$ 13  & 95$\pm$ 4    \\
10 - 14 & 2.42 - 5.43 &     -      & 95 $\pm$ 4 & 112 $\pm$ 6  &     -        \\
15 - 19 & 2.70 - 4.56 & 49 $\pm$ 6 & 82 $\pm$ 3 & 132 $\pm$ 17 & 185 $\pm$ 98 \\
20 - 24 & 1.94 - 4.36 &     -      &      -     & 115 $\pm$ 14 &     -        \\
25 - 29 & 1.66 - 3.73 &     -    & 100 $\pm$ 10 & 122 $\pm$ 14 & 98 $\pm$ 7   \\
30 - 34 & 1.83 - 2.63 &     -      &      -     & 116 $\pm$ 10 & 165 $\pm$ 57 \\
35 - 39 &      -      &     -      &     -      &      -       &     -        \\
40 - 44 & 2.01 - 2.89 & 64 $\pm$ 3 & 86 $\pm$ 3 &      -       &     -        \\
45 - 49 & 3.02 - 4.35 &      -     & 72 $\pm$ 4 & 95 $\pm$ 7   &     -        \\
\hline
\end{tabular}
\end{center}
\end{table}


\clearpage


\clearpage

\bibliographystyle{apj}






\end{document}